# The Effects of the Misfit Structure on Thermoelectric Properties of $Bi_{2-x}Pb_xSr_2Co_2O_y$ Single Crystals

T. Fujii and I. Terasaki
Department of Applied Physics, Waseda University, Tokyo 169-8555, Japan
Precursory Research for Embryonic Science and Technology,
Japan Science and Technology Corporation, Kawaguchi 332-0012, Japan
fujii@htsc.sci.waseda.ac.jp

**Abstract**

In-plane anisotropy of the resistivity and thermopower was measured for single crystal $Bi_{2-x}Pb_xSr_2Co_2O_y$. There is large in-plane anisotropy, which is attributed to the anisotropic pseudogap formation due to the different crystal symmetry between the square $Bi_2Sr_2O_4$ layer and the triangular $CuO_2$ layer. The magnitude of the thermopower both along $a$- and $b$-axis direction increases with Pb doping from x=0 to 0.4, where we observe discontinuous shrink of $b$-axis length. We attribute this to the enhancement of the misfitness. Thus, we can improve the thermoelectric properties by tuning the misfitness.

**Introduction**

Since $NaCo_2O_4$ was found to show large thermopower (100 μV/K at 300 K) and low resistivity (200 μΩcm at 300 K) [1], the layered Co oxides are expected to be a candidate for thermoelectric materials [2-5]. They are characterized by a two-dimensional $CoO_2$ layer, which consists of edge-shared $CoO_6$ octahedra responsible for the electric conduction and the large thermopower. Among them, $Bi_2Sr_2Co_2O_y$ was first thought to have a structure isomorphic to that of the high-$T_c$ Cu oxide $Bi_2Sr_2CaCu_2O_{8+\delta}$ (Bi-2212) because these two compounds have almost same lattice constants. However, it turned out to be a "misfit compound", where the rock-salt type square $Bi_2Sr_2O_4$ layer lies on the $CdI_2$-type triangular $CoO_2$ layer with lattice misfit along the $b$ axis [6]. Another striking feature of the crystal structure in the Bi based layered oxides, is modulation structure [7], which disappears by the substitution of Bi by Pb [8].

From the viewpoint of the group theory, the hexagonal $CoO_2$ layer in $Bi_2Sr_2Co_2O_y$ and the square $CuO_2$ layer in Bi-2212 themselves should have no in-plane anisotropy in the thermopower and resistivity. However, in some cases, in-plane anisotropy in the resistivity can be observed [9]. A possible origin of the in-plane anisotropy is the self-organization of the electronic system into quasi-one-dimensional stripes. The enhancement of the in-plane anisotropy of the resistivity $\rho_a/\rho_b$ has been reported in the lightly doped high-$T_c$ Cu oxide $La_{2-x}Sr_xCuO_4$ and $YBa_2Cu_3O_y$, and has been associated with the charge-stripes [10].

In the case of the $Bi_2Sr_2Co_2O_y$, the coupling between the square $Bi_2Sr_2O_4$ layer and the triangular $CoO_2$ layer is considered to be strong. Then, the stack of different symmetry layers would induce the in-plane anisotropy in the physical properties. Moreover, the lattice misfit between these two layers is thought to induce chemical pressure along the $b$-axis direction. It is, then, interesting how the electric properties are affected by these misfit structures. Here, we have grown the large single crystal of $Bi_{2-x}Pb_xSr_2Co_2O_y$ by using traveling solvent floating zone (TSFZ) method, and have measured the in-plane anisotropy on the resistivity and thermopower.

**Experimental**

Single crystals of $Bi_{2-x}Pb_xSr_2Co_2O_y$ (x=0, 0.4, and 0.6) were grown by TSFZ method at a growth rate of 0.5 mm/h in a mixed gas flow of $O_2$ (20%) and Ar (80%). The starting compositions for all Pb concentrations were applied (Bi,Pb) : Sr : Co = 2 : 2 : 2. Large single crystals with dimensions up to $5*4*0.5$ mm$^3$ were successfully grown.

The grown crystals were characterized by a four-circle x-ray diffractometer (Cu $K_a$ x-ray source) and a transmission electron microscope (TEM). The actual composition was analyzed through inductively coupled plasma-atomic emission spectroscopy (ICP) and energy dispersive x-ray analysis (EDX). $a$- and $b$-axis resistivities were measured by a standard four-probe method using different samples cleaved along $a$- and $b$-axis directions. $a$- and $b$-axis thermopowers were measured by a steady-state technique using the same sample changing the measurement configurations.

**Results and Discussion**

Samples used for this study are listed in Table 1 with the actual chemical compositions. Hereafter, we refer to these samples as nominal Pb concentration. The actual compositions of Pb measured by ICP and EDX correspond well with the nominal compositions. We analyze several samples for each Pb concentration, and confirm the homogeneity of the samples.

Since $Bi_{2-x}Pb_xSr_2Co_2O_y$ has very complicated structure as explained above, it was difficult to perform single-crystal structure analysis. We manually searched the diffraction peak using the reported lattice constants. The obtained lattice parameters are shown in Fig. 1 plotted against actual composition. The data from the previous report are also shown [11], where the lattice constants were estimated from a powder XRD measurement using crushed single crystals. These results well correspond to each other. Since Co atoms have large mass absorption coefficient for Cu $K_a$ radiation, observed reflections come from the rock-salt $Bi_2Sr_2O_4$ layer. Thus, we refer $a$- and $b$-axis lengths obtained here as $a_{RS}$- and $b_{RS}$-axis. The $c$-axis length monotonically increases with increasing Pb concentration due to the large ion radius of Pb. Note that the $b_{RS}$-axis length discontinuously shrink with Pb concentration from x=0 to 0.4. While, the $a_{RS}$-axis length does not change with Pb concentration.

**Table 1.** Chemical composition estimated from energy dispersive x-ray (EDX) analysis and inductively coupled plasma-atomic emission spectroscopy (ICP).

|       |     | Bi   | Pb   | Sr   | Co  |
|-------|-----|------|------|------|-----|
| x=0   | EDX | 2.21 | 0    | 2.20 | 2.0 |
|       | ICP | 2.10 | 0    | 2.10 | 2.0 |
| x=0.4 |     | 1.61 | 0.40 | 2.09 | 2.0 |
|       |     | 1.82 | 0.42 | 2.18 | 2.0 |
| x=0.6 |     | 1.58 | 0.57 | 2.13 | 2.0 |
|       |     | 1.60 | 0.60 | 2.10 | 2.0 |

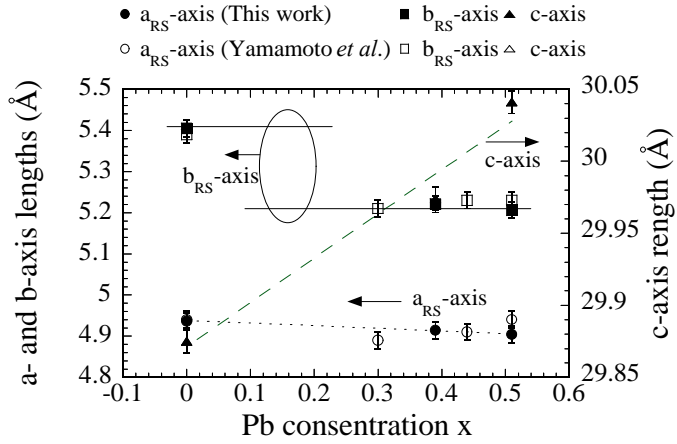

**Figure 1**. The lattice parameter characterized by a four-circle x-ray diffractometer. Previous data estimated from powder x-ray diffractometer using crushed single crystals is also shown.

Figures 2(a) and (b) show the TEM diffraction patterns of $Bi_{2-x}Pb_xSr_2Co_2O_y$ [(a): x=0 and (b): 0.4]. The hexagonal diffraction patterns from the $CoO_2$ layer and the square diffraction patterns from the $Bi_2Sr_2O_4$ layer are clearly observed. The $b$-axis length of the hexagonal $CoO_2$ layer $b_H$ is independent of Pb concentration within the resolution limit, and is about 2.8 Å. From these two diffraction pattern images, we notice that the distance between the $(020_H)$ and $(040_{RS})$ spot of x=0 is narrower than that of x=0.4, indicating the enhancement of misfitness. Another remarkable difference between TEM diffraction patterns is the satellite diffraction patterns along the oblique direction from $a_{RS}^*$ and $b_{RS}^*$ seen in the x=0 sample. This satellite diffraction patterns are due to the modulation structure, which is also observed in high-$T_c$ Cu oxide $Bi_2Sr_2CaCu_2O_{8+\delta}$. The modulation structure was confirmed by the Laue transmission photographs with incident x-ray beam perpendicular to the $ab$-plane. Figure 2(c) and (d) show the Laue patterns of x=0 and 0.4 samples respectively. A two-fold symmetry Laue pattern, where an axis of symmetry is tilted about 45 from $a^*_{RS}$ or $b^*_{RS}$ axis, can be observed, while the Laue pattern of x=0.4 sample shows a four-fold symmetry along the $a^*_{RS}$ and $b^*_{RS}$ axes. This two-fold symmetry Laue pattern agrees well with that of superconducting compound Bi-2212, which have modulation structure [12].

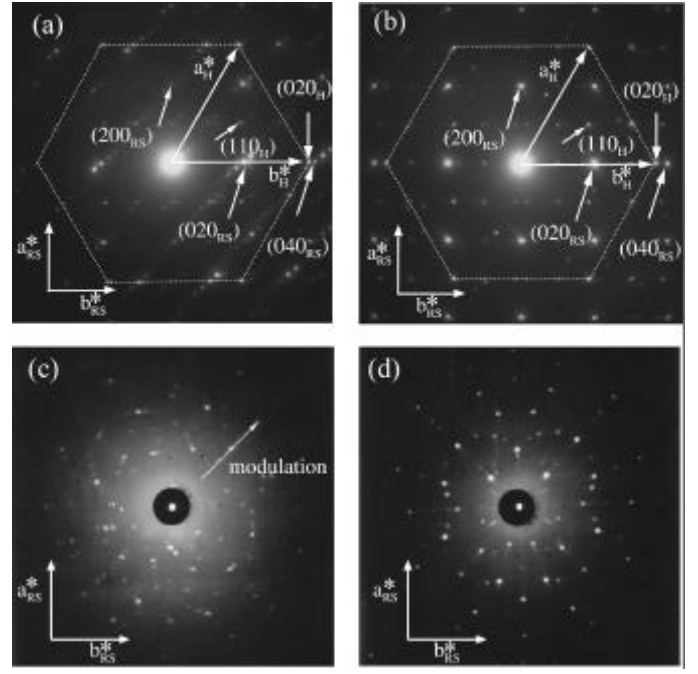

**Figure 2**. TEM diffraction patterns of $Bi_{2-x}Pb_xSr_2Co_2O_y$ [(a): x=0 and (b): 0.4]. Laue transmission photographs with incident x-ray beam perpendicular to the $ab$-plane [(c): x=0 and (d): 0.4]

Figure 3(a) shows the temperature dependence of the $a$- and $b$-axis (of the rock-salt layer) resistivities for various Pb concentrations. The magnitude of the resistivity decreases with increasing Pb concentration, indicating carrier doping by the substitution of divalent Pb for trivalent Bi. These resistivities are normalized at room temperature in Fig. 3(b). The slope of the resistivity increases with Pb concentration, which also indicates the carrier doping. In a rough estimation of the carriers based on the nominal composition, 20% (x=0.4) substitution of Pb introduces 0.2 hole per Co atoms. The decrease of the resistivity from x=0 to 0.4 at room temperature seems to be smaller than expected from the above estimation. Actually, reported Hall coefficient does not change very much, which indicate that less than 0.05 hale per Co atoms is introduced by 20% substitution of Pb [13]. On the other hand, Pb substitution strongly suppresses the upturn behavior at low temperatures. These results suggest that the Pb substitution works not only as carrier doping, but also changes the electronic structure of the $CoO_2$ layer.

Figures 4(a)-(c) show the temperature dependence of the thermopower measured along the $a$- and $b$-axis directions. The magnitude of the thermopower of both directions increases with increasing Pb concentration from x=0 to 0.4. In conventional thermoelectric materials, the thermopower and the resistivity depend on the carrier concentration, and the thermopower is expected to decrease when the resistivity decreases with increasing carrier concentration. Thus the increase in the thermopower is considered to be due to the discontinuous shrink of the rock-salt $b$-axis length, which overcomes the decrease of the thermopower by doping. The chemical pressure induced by the misfit structure enhances the

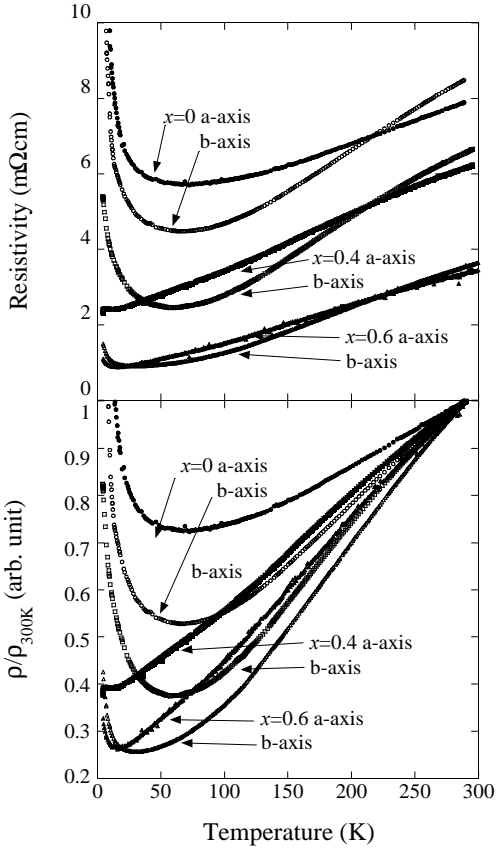

**Figurers 3.** (a) Temperature dependence of the $a$- and $b$-axis resistiviteis. (b) They are normalized at room temperature.

thermopower as is seen in the pressure dependence of the thermopower in the Ce-based compound [14].

On the other hand, the thermopower decreases with further Pb doping from x=0.4 to 0.6. Since the crystal structure is nearly unchanged, observed decrease in the thermopower and resistivity is attributed to the increase of the carrier concentration.

Next we will discuss the in-plane anisotropy in the resistivity ($\rho_b/\rho_a$) and thermopower ($S_b/S_a$). Figures 5(a) and (b) show the temperature dependence of $\rho_b/\rho_a$ and $S_b/S_a$, respectively. For all Pb concentrations, $\rho_b/\rho_a$ decreases with decreasing temperature near the room temperature, while it increase rapidly below 80 K, which indicate that $\rho_a$ is more conductive than $\rho_b$ in low temperature. $\rho_b/\rho_a$ of x=0.4 is as large as 2.5 at 4.5 K, while $\rho_b/\rho_a$ of x=0.6 decreases due to the suppression of the upturn along the $b$-axis. We have previously proposed that the large in-plane anisotropy is come from the anisotropic pseudogap formation. If the misfit structure lowers the crystal symmetry of the $CoO_2$ layer to induce the spin-density-wave-like state, the anisotropic pseudogap would be open [9]. Actually, $S_b/S_a$ and $\rho_b/\rho_a$ are considered to be dominated by the density of states, because the magnitude and the temperature dependence of $S_b/S_a$ and $\rho_b/\rho_a$ roughly correspond to each other.

On the other hand, the increase in the anisotropy at low temperature is quantitatively very similar to lightly doped high-$T_c$ Cu oxides $La_{2-x}Sr_xCuO_4$ (x=0.02-0.04) and $YBa_2Cu_3O_y$ (y=6.35-6.55). These large in-plane anisotropies in high-$T_c$ Cu oxide have been discussed in relation to the stripe order [10]. Since the triangular $CoO_2$ layer itself no in-plane anisotropy, large in-plane anisotropy in the resistivity suggests the self-organization of the electronic system as high-$T_c$ Cu oxide.

By looking carefully, one can find that $\rho_b/\rho_a$ of x=0 is somewhat smaller than $S_b/S_a$. We attribute this to the modulation structure, which works as an anisotropic scattering center in Bi-2212 [15]. The anisotropy would be averaged by the modulation structure, whose direction is tilted by 45 degrees from the $a$- and $b$-axis direction.

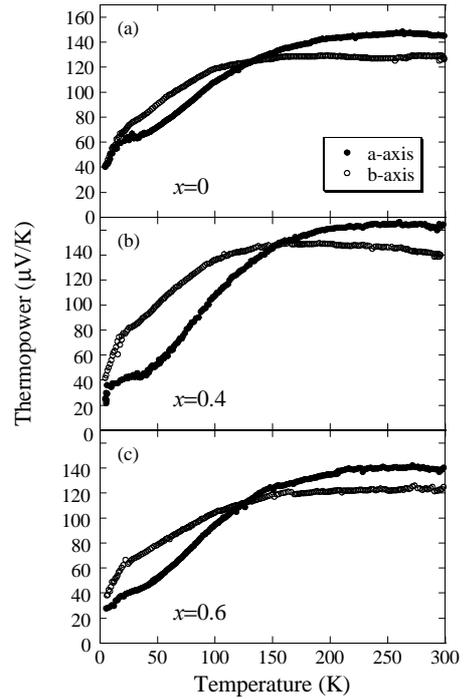

**Figure 4.** Temperature dependence of the thermopower along $a$- and $b$-axes.

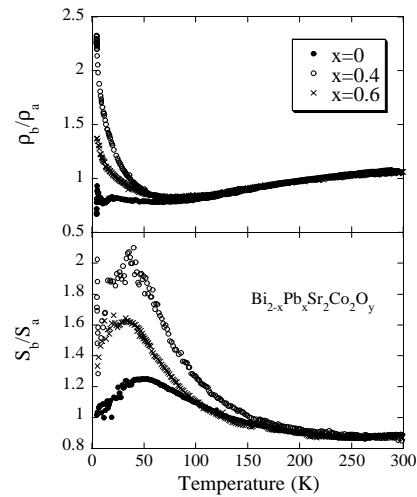

**Figure 5.** In plane anisotropy in the resistivity ($\rho_b/\rho_a$) and thermopower ($S_b/S_a$).

## Conclusions

We have successfully grown the large single crystal of $Bi_{2-x}Pb_xSr_2Co_2O_y$ by TSFZ method and measured the in-plane anisotropy in the resistivity and thermopower. From the structural analysis, significant change in the $b_{RS}$-axis length of the rock salt $Bi_2Sr_2O_4$ layer has observed between x=0 and 0.4, which causes the enhancement of the misfitness. The 15% increase in thermopower is due to the increase of the chemical pressure, as seen in the Ce-based compounds. These results indicate the thermoelectric properties can be controlled by controlling the misfitness. There is a large in-plane anisotropy in the resistivity and thermopower. The anisotropic pseudogap formation at low temperatures would make the resistivity nonmetallic and the thermopower larger along the $b$-axis direction. Another scenario of the large in-plane anisotropy is a charge stripe as seen in the lightly doped high-$T_c$ superconductor. In the case of the resistivity for x=0, the modulation structure along the oblique direction from the $a_{RS}$- and $b_{RS}$-axis averages the anisotropy.

## Acknowledgments

I would like to thank A. Matsuda and T. Watanabe for collaboration at the early stage of this work and also appreciate T. Goto and S. Enomoto for crystal characterization.